\documentstyle[12pt]{article}
\def\ssl#1{\rlap{\hbox{$\mskip 3 mu /$}}#1}

\newcommand{\ft}[2]{{\textstyle\frac{#1}{#2}}}                                   
\newcommand{\eqn}[1]{(\ref{#1})}

\begin{document}


\def\Nequalstwo{\Psi}
\def\eff{{\rm eff}}
\def\inst{{\rm inst}}
\def\fermi{{\rm fermi}}
\def\trtwo{\tr^{}_2\,}
\def\finv{f^{-1}}
\def\Ubar{\bar U}
\def\wbar{\bar w}
\def\fbar{\bar f}
\def\abar{\bar a}
\def\bbar{\bar b}
\def\Deltabar{\bar\Delta}
\def\dalpha{{\dot\alpha}}
\def\dbeta{{\dot\beta}}
\def\dgamma{{\dot\gamma}}
\def\ddelta{{\dot\delta}}
\def\Sbar{\bar S}
\def\Im{{\rm Im}}
\def\sst{\scriptscriptstyle}
\def\cld{C_{\sst\rm LD}^{}}
\def\csd{C_{\sst\rm SD}^{}}
\def\bigI{{\rm I}_{\sst 3\rm D}}
\def\Mr{{\rm M}_{\sst R}}
\def\cJ{C_{\sst J}}
\def\one{{\sst(1)}}
\def\two{{\sst(2)}}
\def\vsd{v^{\sst\rm SD}}
\def\vasd{v^{\sst\rm ASD}}
\def\Phibar{\bar\Phi}
\def\F{{\cal F}_{\sst\rm SW}}
\def\P{{\cal P}}
\def\A{{\cal A}}
\def\susy{supersymmetry}
\def\sigmabar{\bar\sigma}
\def\barsigma{\sigmabar}
\def\ASD{{\scriptscriptstyle\rm ASD}}
\def\cl{{\,\rm cl}}
\def\lambdabar{\bar\lambda}
\def\R{{R}}
\def\psibar{\bar\psi}
\def\sqrtwo{\sqrt{2}\,}
\def\etabar{\bar\eta}
\def\Thetabar{{\bar\Theta_0}}
\def\Qbar{\bar Q}
\def\susic{supersymmetric}
\def\vhiggs{{\rm v}}
\def\vhiggsa{{\cal A}_{\sst00}}
\def\vbarhiggs{\bar{\rm v}}
\def\vhiggsbar{\bar{\rm v}}
\def\novetal{Novikov et al.}
\def\Novetal{Novikov et al.}
\def\ADS{Affleck, Dine and Seiberg}
\def\ads{Affleck, Dine and Seiberg}
\def\setI{\{{\cal I}\}}
\def\Abar{A^\dagger}
\def\B{{\cal B}}
\def\infinity{\infty}
\def\C{{\cal C}}
\def\Psitwo{\Psi_{\scriptscriptstyle N=2}}
\def\Psibartwo{\bar\Psi_{\scriptscriptstyle N=2}}
\def\zero{{\scriptscriptstyle(0)}}
\def\new{{\scriptscriptstyle\rm new}}
\def\u{\underline}
\def\uA{\,\lower 1.2ex\hbox{$\sim$}\mkern-13.5mu A}
\def\uBmu{\,\lower 1.2ex\hbox{$\sim$}\mkern-13.5mu B_\mu}
\def\uAmu{\,\lower 1.2ex\hbox{$\sim$}\mkern-13.5mu A_\mu}
\def\uX{\,\lower 1.2ex\hbox{$\sim$}\mkern-13.5mu X}
\def\uD{\,\lower 1.2ex\hbox{$\sim$}\mkern-13.5mu {\rm D}}
\def\uDzero{{\uD}^\zero}
\def\uAzero{{\uA}^\zero}
\def\upsizero{{\upsi}^\zero}
\def\uF{\,\lower 1.2ex\hbox{$\sim$}\mkern-13.5mu F}
\def\uW{\,\lower 1.2ex\hbox{$\sim$}\mkern-13.5mu W}
\def\uWbar{\,\lower 1.2ex\hbox{$\sim$}\mkern-13.5mu {\overline W}}
\def\Dbar{D^\dagger}
\def\Fbar{F^\dagger}
\def\uAbar{{\uA}^\dagger}
\def\uAbarzero{{\uA}^{\dagger\zero}}
\def\uDbar{{\uD}^\dagger}
\def\uDbarzero{{\uD}^{\dagger\zero}}
\def\uFbar{{\uF}^\dagger}
\def\uFbarzero{{\uF}^{\dagger\zero}}
\def\uV{\,\lower 1.2ex\hbox{$\sim$}\mkern-13.5mu V}
\def\uZ{\,\lower 1.2ex\hbox{$\sim$}\mkern-13.5mu Z}
\def\uv{\lower 1.0ex\hbox{$\scriptstyle\sim$}\mkern-11.0mu v}
\def\uc{\lower 1.0ex\hbox{$\scriptstyle\sim$}\mkern-11.0mu c}
\def\uB{\lower 1.0ex\hbox{$\scriptstyle\sim$}\mkern-11.0mu B}
\def\uPsi{\,\lower 1.2ex\hbox{$\sim$}\mkern-13.5mu \Psi}
\def\uPhi{\,\lower 1.2ex\hbox{$\sim$}\mkern-13.5mu \Phi}
\def\uchi{\lower 1.5ex\hbox{$\sim$}\mkern-13.5mu \chi}
\def\utheta{\lower 1.5ex\hbox{$\sim$}\mkern-13.5mu \theta}
\def\chitilde{\tilde \chi}
\def\etatilde{\tilde \eta}
\def\uchitilde{\lower 1.5ex\hbox{$\sim$}\mkern-13.5mu \tilde\chi}
\def\ueta{\lower 1.5ex\hbox{$\sim$}\mkern-13.5mu \eta}
\def\uetatilde{\lower 1.5ex\hbox{$\sim$}\mkern-13.5mu \tilde\eta}
\def\Psibar{\bar\Psi}
\def\uPsibar{\,\lower 1.2ex\hbox{$\sim$}\mkern-13.5mu \Psibar}
\def\upsi{\,\lower 1.5ex\hbox{$\sim$}\mkern-13.5mu \psi}
\def\uphi{\lower 1.5ex\hbox{$\sim$}\mkern-13.5mu \phi}
\def\uphione{\lower 1.5ex\hbox{$\sim$}\mkern-13.5mu \phi_1}
\def\uph2{\lower 1.5ex\hbox{$\sim$}\mkern-13.5mu \phi_2}
\def\psibar{\bar\psi}
\def\upsibar{\,\lower 1.5ex\hbox{$\sim$}\mkern-13.5mu \psibar}
\def\etabar{\bar\eta}
\def\uetabar{\,\lower 1.5ex\hbox{$\sim$}\mkern-13.5mu \etabar}
\def\chibar{\bar\chi}
\def\uchibar{\,\lower 1.5ex\hbox{$\sim$}\mkern-13.5mu \chibar}
\def\upsibarzero{\,\lower 1.5ex\hbox{$\sim$}\mkern-13.5mu \psibar^\zero}
\def\ulambda{\,\lower 1.2ex\hbox{$\sim$}\mkern-13.5mu \lambda}
\def\ulambdabar{\,\lower 1.2ex\hbox{$\sim$}\mkern-13.5mu \lambdabar}
\def\ulambdabarzero{\,\lower 1.2ex\hbox{$\sim$}\mkern-13.5mu \lambdabar^\zero}
\def\ulambdabarnew{\,\lower 1.2ex\hbox{$\sim$}\mkern-13.5mu \lambdabar^\new}
\def\D{{\cal D}}
\def\M{{\cal M}}
\def\N{{\cal N}}
\def\Dslash{\,\,{\raise.15ex\hbox{/}\mkern-12mu D}}
\def\Dbarslash{\,\,{\raise.15ex\hbox{/}\mkern-12mu {\bar D}}}
\def\delslash{\,\,{\raise.15ex\hbox{/}\mkern-9mu \partial}}
\def\delbarslash{\,\,{\raise.15ex\hbox{/}\mkern-9mu {\bar\partial}}}
\def\L{{\cal L}}
\def\hf{{\textstyle{1\over2}}}
\def\quarter{{\textstyle{1\over4}}}
\def\twe{{\textstyle{1\over12}}}
\def\eighth{{\textstyle{1\over8}}}
\def\fourth{\quarter}
\def\wb{Wess and Bagger}
\def\xibar{\bar\xi}
\def\ss{{\scriptscriptstyle\rm ss}}
\def\sc{{\scriptscriptstyle\rm sc}}
\def\uvcl{{\uv}^\cl}
\def\uAcl{\,\lower 1.2ex\hbox{$\sim$}\mkern-13.5mu A^{}_{\cl}}
\def\uAbarcl{\,\lower 1.2ex\hbox{$\sim$}\mkern-13.5mu A_{\cl}^\dagger}
\def\upsinew{{\upsi}^\new}
\def\ASDzero{{{\scriptscriptstyle\rm ASD}\zero}}
\def\SDzero{{{\scriptscriptstyle\rm SD}\zero}}
\def\SD{{\scriptscriptstyle\rm SD}}
\def\varthetabar{{\bar\vartheta}}
\def\three{{\scriptscriptstyle(3)}}
\def\dagthree{{\dagger\scriptscriptstyle(3)}}
\def\ld{{\scriptscriptstyle\rm LD}}
\def\vld{v^\ld}
\def\Dld{{\rm D}^\ld}
\def\Fld{F^\ld}
\def\Ald{A^\ld}
\def\Fbarld{F^{\dagger\scriptscriptstyle\rm LD}}
\def\Abarld{A^{\dagger\scriptscriptstyle \rm LD}}
\def\lambdald{\lambda^\ld}
\def\lambdabarld{\bar\lambda^\ld}
\def\psild{\psi^\ld}
\def\psibarld{\bar\psi^\ld}
\def\dsiginst{d\sigma_{\scriptscriptstyle\rm inst}}
\def\xione{\xi_1}
\def\xionebar{\bar\xi_1}
\def\xitwo{\xi_2}
\def\xitwobar{\bar\xi_2}
\def\thetatwo{\vartheta_2}
\def\thetatwobar{\bar\vartheta_2}
\def\Ltwo{\L_{\sst SU(2)}}
\def\Leff{\L_{\rm eff}}
\def\Laux{\L_{\rm aux}}
\def\oneloop{{\sst\rm 1\hbox{-}\sst\rm loop}}
\def\LSUtwo{{\cal L}_{\rm SU(2)}}
\def\Dhat{\hat\D}
\def\bkgd{{\sst\rm bkgd}}
\def\Lgft{{\cal L}_{\sst\rm g.f.t.}}
\def\Lghost{{\cal L}_{\sst\rm ghost}}
\def\Sinst{S_{\rm inst}}
\def\etal{{\rm et al.}}
\def\S{{\cal S}}
\def\L{ {\cal L}}
\def\C{ {\cal C}}
\def\N{ {\cal N}}
\def\calE{{\cal E}}
\def\lin{{\rm lin}}
\def\Tr{{\rm Tr}}
\def\mxth{\mathsurround=0pt }
\def\xversim#1#2{\lower2.pt\vbox{\baselineskip0pt \lineskip-.5pt
x  \ialign{$\mxth#1\hfil##\hfil$\crcr#2\crcr\sim\crcr}}}
\def\simgr{\mathrel{\mathpalette\xversim >}}
\def\simle{\mathrel{\mathpalette\xversim <}}
\def\slash{\llap /}
\def\lagr{{\cal L}}

\renewcommand{\a}{\alpha}
\renewcommand{\b}{\beta}
\renewcommand{\c}{\gamma}
\renewcommand{\d}{\delta}
\newcommand{\pa}{\partial}
\newcommand{\g}{\gamma}
\newcommand{\G}{\Gamma}
\newcommand{\e}{\epsilon}
\newcommand{\z}{\zeta}
\newcommand{\Z}{\Zeta}
\newcommand{\K}{\Kappa}
\renewcommand{\l}{\lambda}
\renewcommand{\L}{\Lambda}
\newcommand{\m}{\mu}
\newcommand{\n}{\nu}
\newcommand{\X}{\Chi}

\newcommand{\s}{\sigma}
\renewcommand{\S}{\Sigma}
\renewcommand{\t}{\tau}
\newcommand{\T}{\Tau}
\newcommand{\y}{\upsilon}
\newcommand{\Y}{\upsilon}
\renewcommand{\o}{\omega}
\newcommand{\q}{\theta}
\newcommand{\h}{\eta}
\newcommand{\cmap}{{$\bf c$} map}
\newcommand{\Ka}{K\"ahler} 
\renewcommand{\O}{{\Omega}}
\newcommand{\var}{\varepsilon}
%

\newcommand{\nd}[1]{/\hspace{-0.5em} #1}
\begin{titlepage}
\begin{flushright}
UW/PT 98-1 \\
SWAT-98/184 \\
hep-th/9803065
\end{flushright}
\begin{centering}
\vspace{.2in}
{\large {\bf Instanton Effects in Three-Dimensional Supersymmetric \\
Gauge Theories with Matter}}\\
\vspace{.4in}
 N. Dorey$^{1,2}$,  D. Tong$^{1,2}$  and 
S. Vandoren$^{2}$\\
\vspace{.4in}
$^{1}$ Department of Physics, University of Washington, Box 351560   \\
Seattle, Washington 98195-1560, USA\\
\vspace{.3in}
$^{2}$ Department of Physics, University of Wales, Swansea \\
Singleton Park, Swansea, SA2 8PP, UK\\
\vspace{.4in}
{\bf Abstract} \\
\end{centering}
Using standard field theory techniques we compute perturbative and 
instanton contributions to the Coulomb branch of three-dimensional 
supersymmetric QCD with $N=2$ and $N=4$ supersymmetry and gauge 
group $SU(2)$. For the $N=4$ theory with one massless flavor, 
we confirm the proposal of Seiberg and Witten that the Coulomb branch 
is the double-cover of the centered moduli space of two BPS monopoles 
constructed by Atiyah and Hitchin. Introducing a hypermultiplet 
mass term, we show that the asymptotic metric on the 
Coulomb branch coincides with the metric on Dancer's deformation of the 
monopole moduli space. For the $N=2$ theory with $N_f$ flavors, we compute the 
one-loop corrections to the metric and complex structure on the Coulomb branch.
We then determine the superpotential including one-loop effects around the 
instanton background. These calculations provide an explicit check 
of several results previously 
obtained by symmetry and holomorphy arguments.


\end{titlepage}

\section{Introduction}
\paragraph{}
 
Recently several exact results have been proposed for supersymmetric 
gauge theories in three dimensions (3D). The Coulomb branches of $N=4$
theories (theories with eight supercharges) 
were examined in \cite{sw3,chahan} from a field theory perspective where a 
connection to monopole moduli spaces was revealed. 
The arguments presented in these papers rely on certain assumptions about 
the strong coupling behaviour of these theories which are motivated by input 
from  string theory \cite{seib}. These theories were further studied using 
D-brane technology \cite{hanwit} where an alternative
derivation of these results was presented. 
For $N=4$ supersymmetric Yang-Mills theory (SYM) without matter, 
Seiberg and Witten (SW) proposed \cite{sw3}
that the Coulomb branch of the theory is 
the centered moduli space of two BPS monopoles constructed 
by Atiyah and Hitchin (AH) \cite{AH}.
This correspondence was confirmed in \cite{dkmtv} by performing explicit
one-loop calculations in the vacuum and one-instanton sectors. In fact,
as explained in \cite{dkmtv}, the symmetries of the $N=4$ theory are so
restrictive that the exact form of the metric on the Coulomb branch is
uniquely determined by this weak-coupling data. In this sense the
calculations presented in \cite{dkmtv} can be considered as a direct
proof of the SW proposal.

In the present paper we will generalise these calculations to $N=4$ theories
coupled to matter. In particular, for the theory with one massless flavor, SW
have proposed that the Coulomb branch is the double-cover 
of the AH manifold. In the following we 
confirm this by performing a two-instanton calculation. SW
further claimed that the mass parameter of a single hypermultiplet 
plays the role of Dancer's deformation parameter of this manifold 
\cite{Dancer}. We show that, at least in perturbation theory, 
this is indeed the case. 

Theories with $N=2$ supersymmetry (four supercharges) 
dynamically generate a superpotential 
on the Coulomb branch. This was calculated for the pure $SU(2)$ theory 
many years ago in \cite{3dN=2}. More recently, proposals for the 
exact superpotential have been made for a wide range of $N=2$ theories 
\cite{berkN=2,aha3d}. In this paper, we 
calculate the one-instanton contributions to the 
superpotential for $SU(2)$ theories with matter   
and confirm the corresponding predictions of \cite{berkN=2,aha3d}.  
We also compute the one-loop correction
to the \Ka\ metric and complex structure for these models. 
To show explicitly that the instanton-generated
superpotential is holomorphic, an effect discovered in \cite{dkmtv}
turns out to be crucial. In that paper it was shown in the context of
the $N=4$ theory, that the one-loop determinants arising from
integration over bose and fermi fluctuations in the instanton background
do not cancel exactly. In the following we will show that, in the $N=2$
theory, the corresponding residual factor precisely reflects the fact
that the superpotential is holomorphic with respect to the one-loop
corrected complex structure. We further show how the renormalization 
group decoupling of massive hypermultiplets is manifest, both in perturbation
theory and in the 
instanton calculus. This enables us to flow from the 
$N=4$ to the $N=2$ model, or from $N_f$
to $N_f-1$ flavors, by taking an infinite mass limit.

The paper is organised as follows. In section 2 we discuss classical 
aspects of the models we consider, paying special attention to the 
different possible mass terms and global symmetries. Section 3 is devoted to 
perturbative effects. In particular, we show that the 
perturbative metric on the Coulomb branch of $N=4$ SYM with a single 
massive hypermultiplet is indeed given by the asymptotic metric 
for the Dancer spaces, in agreement 
with \cite{sw3}. For the $N=2$ model, we determine the one-loop 
correction to the \Ka\ metric and the complex structure.
Some details of these perturbative calculation are presented in an appendix. 
In section 4 we develop the instanton calculus in theories 
with arbitrary matter content, discussing the zero mode structure, the 
collective coordinate measure and the one-loop correction which comes from 
Gaussian integration over fluctuations around the 
background of the instanton. In section 5 we restrict ourselves to the 
$N=4$ theory. For a single massless hypermultiplet 
in the fundamental representation, we confirm that the Coulomb branch is 
the double cover of the AH space. This involves an integration 
over the relative moduli space of two three-dimensional instantons, 
which is itself the AH manifold! Finally, in section 6, we consider the 
$N=2$ model. We compute the superpotential up to one-loop around the 
background of an instanton and compare our calculations with the proposals 
of \cite{3dN=2,berkN=2,aha3d}.

\section{Fields and Symmetries}
\paragraph{}
In this section we discuss the classical theories. 
In the following, $N$ stands for the number of real two-component
Majorana supercharges in 3D theories, whereas 
${\cal N}=N/2$ denotes the number of complex two-component 
supercharges which is the 
usual counting for four dimensional (4D) theories.
Three-dimensional theories with $N=2$ and $N=4$ SUSY can be obtained by
dimensional reduction of the minimal supersymmetric theories theories in 
four and six dimensions respectively. As we discuss below, it is also 
possible to include additional mass terms which have no counterpart in higher 
dimensions. Unless otherwise 
stated, the notation is as in \cite{dkmtv} (e.g. $\uX=X^a\tau^a/2$ denotes 
adjoint-valued fields). 

The multiplets of interest of $N=2$ SUSY in 3D are obtained by dimensional 
reduction of ${\cal N}=1$ gauge and chiral multiplets in 4D. The 3D gauge 
multiplet consists of the gauge field, $\uAmu$, one two-component 
Dirac spinor, $\ulambda  $, and a real scalar field, $\uphi$, which comes 
from the component of the 4D gauge field in the reduced direction. The 
dimensional reduction of the 4D chiral multiplet is the so-called 
`half-hypermultiplet' which consists of a single complex scalar and a single 
Dirac fermion. The $N=2$ SUSY algebra inherits the chiral 
R-symmetry of the four-dimensional ${\cal N}=1$ theory, which we will denote 
$U(1)_{N}$.   

Multiplets of $N=4$ supersymmetry in three dimensions are obtained by 
combining $N=2$ multiplets. The $N=4$ gauge multiplet consists of the $N=2$ 
gauge multiplet introduced above, together with an adjoint half-hypermultiplet 
which includes a complex scalar, $\uA$, and a second 3D Dirac spinor, 
$\upsi$. The action for $N=4$ SYM theory with gauge group $SU(2)$ 
is\footnote{We have denoted
$|[\phi,A]|^2=[\phi,\Abar][\phi,A]$, and fermion multiplication as e.g.
$\l\psi\equiv -i\l^T\g_0\psi$.}
\begin{eqnarray}
S_{VM}&=&\frac{2\pi}{e^2}\int {\rm d}^3x\mbox{ Tr }\{ -\ft12\uv_{\mu\nu}
\uv^{\mu\nu}
+D_\mu \uphi D^\mu \uphi
+2i\ulambdabar \Dslash  \ulambda +2\ulambdabar [\uphi,\ulambda]\nonumber\\
&&\hspace{2cm}+2D_\mu \uAbar D^\mu \uA +2i\upsibar \Dslash \upsi
+[\uA,\uAbar]^2+2|[\uphi,\uA]|^2\nonumber\\
&&\hspace{2cm}+2\sqrt 2i([\uAbar,\upsi]\ulambda+\ulambdabar[\uA,\upsibar])+2\upsibar [\uphi,\upsi]\}\ .\label{SVM}
\end{eqnarray}
Here, the first line taken in isolation is the action of the $N=2$ SYM theory.
The second and third lines provide the gauge couplings, Yukawa couplings and 
potential terms for the adjoint half-hypermutiplet required to complete the 
$N=4$ theory. 

$N=4$ SYM is the dimensional reduction of the 
minimal supersymmetric theory in six dimensions. 
The action \eqn{SVM} has an $SU(2)_R
\times SU(2)_N$ global R-symmetry group. The $SU(2)_N$ is an extension of 
the four-dimensional chiral $U(1)_N$, and corresponds to rotations in
the three reduced directions of the six-dimensional (6D) theory.
The $SU(2)_R$ is already present in 
the 6D theory. The bosons $\uA=(\uphione+i\uph2)/\sqrt 2$ and 
$\uphi=\uphi_{3}$ combine to form a 
triplet under $SU(2)_N$ which we denote\footnote{In the following, the vector 
notation $\vec{X}=(X_{1},X_{2},X_{3})$ always denotes a 
vector of $SU(2)_{N}$}  $\vec{\uphi}=(\uphi_{1},\uphi_{2},\uphi_{3})$. 
These scalars are singlets of $SU(2)_R$. 
The fermions $\ulambda$ and $\upsi$ form a doublet under $SU(2)_R$.
They also transform under $SU(2)_N$ (see e.g. \cite{dkmtv}). 
The potential has flat directions, allowing the scalars to acquire a VEV. 
By an $SU(2)_N$ rotation, we can choose the vacuum to be $\langle\uA\rangle=0, 
\langle\uphi\rangle={\sqrt 2} \vhiggs \tau^3/2$.  
The resulting moduli space of vacua is known as the Coulomb branch of
the theory. Along these directions the $SU(2)$ gauge group is broken
down to $U(1)$ by the  the adjoint Higgs mechanism. 
Gauge fields of the unbroken $U(1)$ subgroup remain massless, while the
remaining gauge bosons receive a mass
$M_W={\sqrt 2} \vhiggs$. On the Coulomb branch the $SU(2)_{N}$ symmetry is 
broken to an abelian subgroup, which is denoted $U(1)_{N}$ as in \cite{sw3}. 
This is the same $U(1)_N$ as exists in the $N=2$ theory.

One can also add matter couplings to (\ref{SVM}) in a way that 
preserves either the full $N=4$ supersymmetry or an $N=2$ subalgebra. 
A hypermultiplet of $N=4$ SUSY consists of 
two half-hypermultiplets transforming in conjugate representations of the 
gauge group. Although any number of half-hypermultiplets preserves $N=2$
supersymmetry, for complex representations of the gauge group 
the theory suffers a $Z_2$ anomaly and Chern-Simons terms 
are dynamically generated unless the half-hypermultiplets are paired to form 
hypermultiplets. We will not consider the anomalous case. 
In the following we will introduce $N_{f}$ hypermultiplets 
in the fundamental representation of $SU(2)$. 
Each hypermultiplet contains two complex scalars, 
$q_i$ and $\tilde q_i$,  and two Dirac spinors, $\psi_i$ and 
$\tilde\psi_i$, $i=1,..,N_f$. As the fundamental representation of 
$SU(2)$ is pseudo-real, all fields transform in the same representation. 

The hypermultiplet action respecting $N=2$ supersymmetry is given by, 
\begin{eqnarray}
\frac{e^2}{2\pi}S_{HM}&=&\int {\rm d}^3x\,\{D_\mu q_i^\dagger D^\mu q_i+D_\mu \tilde q_i^\dagger D^\mu \tilde q_i
+i\bar \psi_i \Dslash \psi_i +i{\bar {\tilde \psi_i}} \Dslash {\tilde \psi_i} \nonumber\\
&&\hspace{1cm}+i\sqrt 2(q_i^\dagger \ulambda \psi_i-\bar \psi_i \ulambdabar q_i)+i\sqrt 2(\tilde q_i \ulambdabar {\bar {\tilde 
\psi_i}}
-\tilde \psi_i \ulambda \tilde q_i^\dagger) \\
&&\hspace{1cm}+{\bar \psi}_i\uphi \psi_i+{\bar {\tilde \psi}}_i\uphi 
{\tilde \psi}_i -\tilde q_i^\dagger\uphi^2\tilde q_i -q_i^\dagger\uphi^2
q_i -\ft18 (q_i^\dagger\tau^aq_i-\tilde q_i\tau^a\tilde q_i^\dagger)^2  \}\ 
.\nonumber\label{SHM}
\end{eqnarray}
One can also introduce a holomorphic superpotential, 
$W(\uA,q_{i},\tilde{q}_{i})$, while preserving $N=2$ supersymmetry. 
In order to increase the number of supersymmetries to $N=4$, 
we must take $W=\sqrt 2{\tilde q}_i\uA q_i$ with the corresponding action 
given by, 
\begin{eqnarray}
\frac{e^2}{2\pi}S_{W}=\int {\rm d}^3x\{-\ft12\frac{\pa^2 W}
{\pa \phi_B \pa \phi_C}\psi_B\psi_C+\mbox{h.c.}
-\sum_B\left|\frac{\partial W}{\partial\phi_B}\right|^2
+q_i^\dagger [\uA^\dagger ,\uA]q_i-\tilde q_i[\uA^\dagger ,\uA] 
{\tilde q_i}^\dagger \}\ ,\label{SW} 
\end{eqnarray}
where we have defined $\phi_B=\{\uA,q_{i},\tilde q_i\}$ and $\psi_B=
\{\upsi,\psi_{i},\tilde \psi_i\}$ for $B=1,2,3$. The final 
two terms in \eqn{SW} arise from integrating out the auxiliary 
field associated with the 4D vector multiplet (``D-terms''). They 
ensure the bosonic potential is invariant 
under $SU(2)_N$. In this case the scalars $(q,\,{\tilde q}^\dagger)$ 
form a doublet under $SU(2)_R$ and do not transform under $SU(2)_N$. 
The hypermultiplet fermions are singlets under $SU(2)_R$ but do transform 
under $SU(2)_N$, see \cite{djdwkv}.  

Finally, we discuss mass terms. Four-dimensional ${\cal N}=2$ 
theories permit a complex mass parameter, $m$, for each hypermultiplet. 
This is most conveniently described by changing the superpotential 
to $W=\sqrt 2{\tilde q}_i\uA q_i+m\tilde q_i q_i$. 
Three-dimensional $N=4$ theories allow for an extra real mass parameter, 
$\tilde{m}$, corresponding to the mass of 3D Dirac fermions.
Unlike the complex mass, it cannot be written as part of the superpotential. 
However the real mass can be introduced in a manifestly supersymmetric way 
by gauging a subgroup of the global flavor 
symmetry and then freezing the gauge multiplet to a 
background scalar expectation value. 
In a real basis the three masses $\vec m=({\rm Re}(m),{\rm Im}(m),\tilde{m})$ 
transform as a vector under $SU(2)_N$. In component form we have,
\begin{eqnarray}
\frac{e^2}{2\pi}S_m&=&\int {\rm d}^3x\,\{-m_i\psi_i\tilde \psi_i-\bar m_i
\bar \psi_i {\bar {\tilde \psi_i}}-\tilde{m}(\bar \psi_i \psi_i+
{\bar {\tilde \psi}}_i {\tilde \psi_i})\nonumber \\ 
&&\hspace{1cm}-|\vec{m}|^2(\tilde q_i\tilde q_i^\dagger+q_iq_i^\dagger)-
\sqrt{2}(\tilde q_i{\vec m}\cdot{\vec {\uphi}}\tilde q_i^\dagger+q_i^\dagger 
{\vec m}\cdot{\vec{\uphi}}q^i)\}
\ .\label{Sm}
\end{eqnarray}
We will also consider the mass deformed $N=4$ obtained by giving 
a mass, ${\vec M}$, to the adjoint half-hypermultiplet
\begin{equation}
\frac{e^2}{2\pi}S_M=\mbox{ Tr }\int {\rm d}^3x \{-M\upsi \upsi-\bar M\upsibar 
\upsibar-\tilde{M}\upsibar \upsi-|{\vec M}|^2\uA^\dagger \uA \}
\ .\label{mdeform}
\end{equation} 
Taking an infinite mass limit for the adjoint half-hypermultiplet,  
the theory flows to $N=2$ SYM. Similarly, on taking the mass of a 
single fundamental hypermultiplet to infinity, the theory with $N_{f}$ 
flavors flows to the theory with $N_{f}-1$ flavors. 
This decoupling provides a useful check on the perturbative and 
instanton calculations presented below. 

\section{Perturbation Theory and the Effective Action}
\paragraph{}
In three dimensions the gauge coupling has the dimensions of mass and 3D 
gauge theories are therefore super-renormalisable. The 
perturbation series is organised in powers of the dimensionless 
quantity $e^2/M_W$. 
In the following, we will consider the finite renormalisation of 
the coupling constant at one-loop. This may be calculated by integrating out 
high frequency modes to obtain the Wilsonian effective action for the 
massless fields.  An explicit calculation of this effect for the $N=4$ theory 
without matter was presented in \cite{dkmtv} 
(In particular, see Appendix B of this reference.). In this section, 
we present the generalization of these results to include additional matter
multiplets with and without mass terms.  We then use the one-loop 
renormalization of the coupling to determine 
the asymptotic behaviour of the metric on the Coulomb branch. 
We initially restrict ourselves to $N=4$ model. The case of $N=4$ with one 
massive flavor will be discussed in detail. 
The $N=2$ case is treated at the end of the section. 

Consider massless hypermultiplets, which can be in the adjoint or 
fundamental representations of the gauge group. 
Let $N_a$ and $N_f$ denote the number of massless 
hypermultiplets in these representations respectively.
A straightforward modification of the calculation given in \cite{dkmtv} 
shows that the one-loop renormalisation to the coupling constant is given by 
the replacement    
\begin{equation}
\frac{2\pi}{e^2}\rightarrow \frac{2\pi}{e^2}\left(1-\frac{2-N_f-2N_a}
{S_{\rm cl}}\right)\ ,
\label{TN1}
\end{equation}
where, for later convenience, we have expressed the answer in terms of 
$S_{\rm cl}\equiv 8\pi^2M_W/e^2$ which is equal to the instanton action. 
Some details of this result and its generalization to include 
hypermultiplet masses are given in the appendix. Here we 
concentrate on the case of one massive fundamental hypermultiplet. 
The corresponding renormalization is   
\begin{equation}
\frac{2\pi}{e^2}\rightarrow \frac{2\pi}{e^2}\left(1-\frac{2}{S_{\rm cl}}
+\frac{e^2}{2^5\pi^2}\sum_{\epsilon =1,2}|\vec{m}+(-1)^\epsilon 
\vec {\vhiggs}/\sqrt{2}|^{-1}\right)\ .
\label{Dancerpert}
\end{equation}
The above expression correctly goes over to the  
$N_{f}=0$ and $N_{f}=1$ results in 
the decoupling limit $|\vec{m}|\rightarrow\infty$ and the massless limit
$|\vec{m}|\rightarrow 0$ respectively. 

To see how these results determine the one-loop metric on the Coulomb branch, 
we must first perform a duality transformation, eliminating 
the massless photon in favour of a periodic scalar. 
We follow closely the discussion 
and notation in \cite{dkmtv}.  
At the classical level, the bosonic low-energy action is simply given by 
the free massless expression
\begin{equation}
S_{\rm B}=\frac{2\pi}{e^{2}}
\int\,{\rm d}^{3}x \left[
\quarter v_{\mu \nu}v^{\mu \nu}+\hf\partial_{\mu}{\vec \phi}\cdot
\partial^{\mu}{\vec \phi}\right]\ ,
\end{equation}
where $v_{\mu\nu}={\rm Tr}(\uv_{\mu\nu}\tau^3)$  with similar definitions 
for the massless scalar fields. 
In three dimensions, the photon is dual to a scalar that 
serves as a Lagrange multiplier for the Bianchi identity. Hence we add to 
the action a term,  
\begin{equation}
S_{\rm S}=\frac{i}{8\pi}\int\,{\rm d}^{3}x\,\s \varepsilon^{\mu\nu\rho}\partial_{\mu}v_{\nu\rho} \ .
\label{surface}
\end{equation}
In a topologically trivial background we may integrate this term by parts, 
disregarding the surface term, to find that the action only depends on $\s$ 
through its derivatives and therefore the theory has a trivial symmetry $\sigma
\rightarrow \sigma+ c$ for constant $c$. On the other hand, in the presence of 
an instanton, the surface term is non-zero.  
In fact, the normalization of (\ref{surface}) was chosen so 
that $S_{\rm S}=-i\s$ in the background of an instanton 
of unit magnetic charge. It follows that 
$\s$ is a periodic variable with period $2\pi$. 
Integrating out the abelian field strength 
then yields the bosonic effective action,
\begin{equation}
S_{\rm B}=\frac{2\pi}{e^{2}}
\int\,{\rm d}^{3}x \ \hf\partial_{\mu}{\vec \phi}\cdot
\partial^{\mu}{\vec \phi} +\frac{2e^{2}}{\pi(8\pi)^{2}} \int\,{\rm d}^{3}x \ 
\hf\partial_{\mu}\sigma\partial^{\mu}\sigma \ .  
\label{lcl}
\end{equation}
At one-loop, the above action is modified by the replacement 
(\ref{TN1}). Although there are other one-loop effects, as we explain below 
the full effective action at one-loop is uniquely determined by the coupling 
constant renormalization.      

On the Coulomb branch, the massless 
fields consist of four real scalar fields and two Dirac fermions, which 
can be seen as four real Majorana fermions. The most general form for the 
low-energy effective action up to two derivatives and four 
fermi terms takes the form of a supersymmetric sigma model with a 
four dimensional target manifold, ${\cal M}$. Due to $N=4$ supersymmetry, the
metric on ${\cal M}$ is hyper-\Ka\ \cite{AG}. In some set of real coordinates, 
$X^a$, and their supersymmetric Majorana partners, $\Omega^a$, $a=1,..,4$, the 
action reads
\begin{equation}
S_{\rm eff}=K \int\,{\rm d}^{3}x \ \left\{\hf g_{ab}(X) \ 
\left[\partial_{\m}X^{a}\partial^\m X^{b}
+{\bar \Omega}^{a}\ssl{D}\Omega^{b}\right] +
\twe R_{abcd}({\bar \Omega}^{a}\Omega^{c})({\bar \Omega}^{b}\Omega^{d})
\right\}\ ,
\label{seff1}
\end{equation}
where the derivative $\ssl{D}$ is covariant 
with respect to the hyper-\Ka\ metric $g_{ab}$, and $R_{abcd}$ denotes the 
Riemann tensor on ${\cal M}$. For later convenience we have included 
an overall normalisation constant $K$ whose value will be fixed below. 

In addition to the hyper-\Ka\ property, the part of the metric 
derived solely from perturbation theory must have a $U(1)$ isometry 
corresponding to the freedom to shift $\s$ by a constant. 
As mentioned above, this symmetry is 
broken by instantons. However it is respected by all perturbative corrections. 
Any four-dimensional hyper-\Ka\ metric 
with such a triholomorphic isometry can be written in the form \cite{Pedpoon},
\begin{equation}
g_{ab}{\rm d}X^a{\rm d}X^{b}=U(\vec{r}){\rm d}\vec{r}\cdot {\rm d}\vec{r}+
4U^{-1}(\vec{r})({\rm d}X^4+\vec{w}\cdot 
{\rm d}\vec{r})^2\ ,
\label{asmetric}
\end{equation}
where $\vec{r}=(X^1,X^2,X^3)$ and $\vec{w}$ satisfies $\vec{\nabla}\times 
\vec{w}=-(1/2)\vec{\nabla}U$. Hence we may determine the metric at one-loop by 
calculating the function $U$ to this order. This may be accomplished 
by comparing the general forms (\ref{seff1}) and (\ref{asmetric}) 
with our classical effective action (\ref{lcl}) suplemented by the 
one-loop replacement (\ref{Dancerpert}).  From this comparison we may deduce 
the relationship between the fields, ${\vec \phi}$ and $\sigma$ and 
the coordinates, $X^a$, on the manifold correct to one loop. 
We fix the normalisation 
by requiring the classical metric to have $U=1$ and, (for $N_f=1$), $X^4=\s$.
This implies the identification, ${\vec r}=(16\pi^2/e^2){\vec \phi}$ and 
$K=2e^2/2^8\pi^3$ and hence the radial coordinate is given as $r=2S_{\cl}$.  
In general these relations will be modified by higher loop corrections.

Using these identifications, the function $U$ is given at one-loop as, 
\begin{equation}
U=1-\frac{4}{r}+\frac{1}{|{\vec \mu}-{\vec r}|}+\frac{1}{|{\vec \mu}+{\vec r}|}
\ ,\label{U}
\end{equation}
where $\vec \mu=(2^5\pi^2/e^2)\vec m$. 
In \cite{sw3}, Seiberg and Witten proposed that the the Coulomb branch of 
the $N=4$ theory with one massive hypermultiplet 
coincides with the three-parameter family of deformations 
${\cal M}(\vec{\l})$ of the double cover of the AH manifold 
discovered by Dancer \cite{Dancer}. The deformation parameter ${\vec{\l}}$ 
was identified as a multiple of the hypermultiplet mass $\vec{m}$. In fact 
the metric on Dancer's manifold is known explicitly only in the 
asymptotic regime. It can be extracted from the $SU(3)$ $(2,1)$ 
monopole moduli space in the limit that the third (distinct) monopole becomes 
infinitely massive \cite{houghton}. It is indeed of the 
form (\ref{asmetric}) with $U$ given by (\ref{U}) and $\vec{\l}=\vec{\mu}$, 
see eqn. (27) in \cite{chalm}. 
This shows that our perturbative calculation is
in agreement with the conjecture of Seiberg and Witten.

Finally, we turn to the $N=2$ model with $N_{f}$ fundamental hypermultiplets. 
The massless fields are a single real scalar $\phi=\Tr(\uphi\tau^{3})$ and the dual photon $\sigma$. 
At the classical level the bosonic low-energy effective action is, 
\begin{equation}
S_{\rm B}=\frac{2\pi}{e^{2}}
\int\,{\rm d}^{3}x \ \hf\partial_{\mu}{\phi}
\partial^{\mu}{\phi} +\frac{2e^{2}}{\pi(8\pi)^{2}} \int\,{\rm d}^{3}x \ 
\hf\partial_{\mu}\sigma\partial^{\mu}\sigma \ .  
\label{lclN=2}
\end{equation}
As in the $N=4$ case, the action is modified at one-loop by a finite 
renormalization of the gauge coupling. The one-loop correction is 
determined by the results given in the Appendix, see \eqn{renccN=2}. 
In the case of massless hypermultiplets,   
\begin{equation}
\label{n=2pert}
\frac{2\pi}{e^2}\rightarrow \frac{2\pi}{e^2}\left(1-\frac{3-N_{f}}{S_{\rm cl}}
\right)\ .\label{pertN=2}
\end{equation}

As before, we would like to compare the result of our 
perturbative calculation with the corresponding terms in the 
most general action allowed by the symmetries of the theory. 
The terms with two derivatives and their supersymmetric completion 
must again take the form of a supersymmetric non-linear sigma model. 
However in the $N=2$ case the target is a 
\Ka\ manifold of complex dimension one.       
In terms of complex coordinates $Z$, $\bar{Z}$ and 3D Dirac  
superpartners $\Psi$, $\bar{\Psi}$ the effective action is, 
\begin{equation}
S_{\rm eff}=L \int\,{\rm d}^3x\,\left\{ g_{{\bar Z}Z}
\left[\partial_\m {\bar Z}\partial ^\m Z  +
{\bar \Psi}\ssl{D}\Psi\right] +\ft14 R(Z,\bar{Z})
{\bar \Psi}^{2}\Psi^{2}\right\}\ .
\label{comp}
\end{equation}
The metric is derived from a \Ka\ potential, 
$g_{\bar{Z}Z}=\partial_{\bar{Z}}\partial_{Z}K(Z,\bar{Z})$, and 
$R(Z,\bar{Z})$ is the non-vanishing component of the Riemann tensor. 

The classical effective action (\ref{lclN=2}) 
can trivially be written in the form (\ref{comp}) with complex 
variable $Z=S_{\rm cl}-i\sigma$ and \Ka\ potential $K=Z\bar{Z}$. This 
normalization fixes the overall 
constant in (\ref{comp}) as $L=2e^2/(\pi(8\pi)^2)$. 
To incorporate the one-loop effect, one must redefine
the complex coordinates as \cite{berkN=2,aha3d}  
\begin{equation}\label{Z}
Z=S_{\rm cl}-(3-N_{f})\log S_{\rm cl}-i\s\ .\label{ZS}
\end{equation}
The one-loop \Ka\ metric is $(1-(3-N_{f})/S_{\rm cl})^{-1}$ which is defined as an 
implicit function of $Z+\bar{Z}$ by (\ref{Z}). 
Using the results of the appendix, 
one easily generalises this to $N_{f}$ massive flavors. 
The result for the one-loop 
corrected complex structure can be written as
\begin{equation}
Z=S_{\rm cl}-i\s-3\log S_{\rm cl}+\ft12 \log\left(
\prod_{i=1}^{N_f}\left[\frac{\sqrt{|m_i|^2+({\tilde m}_{i}+\frac{1}{2}M_W)^2}
+{\tilde m}_{i}+\frac{1}{2}M_W}{\sqrt{|m_i|^2+({\tilde m}_{i}-
\frac{1}{2}M_W)^2}+{\tilde m}_{i}-\frac{1}{2}M_W}\right]\right)\ .
\label{Z1loop}
\end{equation}
Notice that, on taking the masses of one of the $N_f$ flavors to infinity 
we correctly recover the result for $N_{f}-1$ flavors. 

In the $N=2$ case, it is also possible to generate a holomorphic 
superpotential $W(Z)$ 
which preserves supersymmetry. This gives rise to a bosonic 
potential in the low-energy effective action as well as fermion bilinear 
terms; 
\begin{equation}
S_{W}=\int\,{\rm d}^{3}x\, \left\{ \left|\frac{\partial W}
{\partial Z}\right|^{2} + \ft12\frac{\partial^{2}W}
{\partial Z^{2}}\Psi^{2} + \ft12\frac{\partial^{2}\bar{W}}
{\partial \bar{Z}^{2}}\bar{\Psi}^{2}\right\}\ .
\label{superp}
\end{equation}
As mentioned above the perturbative theory has a trivial symmetry under 
which $\sigma$ and therefore $Z$ transforms additively. Clearly this implies 
that the above superpotential cannot be generated at any order 
in perturbation theory. 
However instantons break this symmetry and can contribute terms to $W$ 
such as $\exp(-Z)$ which are periodic in $\sigma$ with period $2\pi$.          
Importantly, the instanton-induced superpotential is holomorphic in the 
superfield $Z$ which implies non-trivial 
quantum corrections in the instanton background when 
re-expressed in terms of $S_{\rm cl}$ and $\sigma$. In Section 5 we will check 
explicitly that the instanton contribution is holomorphic with respect to the 
one-loop corrected complex structure.    

\section{Supersymmetric Instantons in 3D}
\paragraph{} 
The study of instantons in three-dimensional gauge theories 
was initiated by Polyakov \cite{pol} in the
context of quark confinement. The relevant  
field configurations of finite Euclidean action are the static monopole 
solutions of (3+1)-dimensional gauge theory. In the supersymmetric theories 
considered here, the Prasad-Sommerfield limit is automatic and hence the 
instantons we will consider are BPS monopoles.    

 Instanton effects in the three-dimensional gauge theories with $N=4$ 
supersymmetry were studied in  
\cite{sw3,chahan,hanwit}. Explicit computations were performed for the  
$SU(2)$ theory without matter 
in \cite{dkmtv} and our methods here follow the same strategy.  
It is also useful for the reader to consult \cite{dkm3d}.  
In the present section we will discuss the zero modes of the instanton 
and the measure for integration over the collective 
coordinates, as well as the non-cancelling 
determinants coming from integration over 
quadratic fluctuations around the instanton background.
In section 5, we will perform a two instanton calculation in the 
case of one massless 
hypermultiplet, which confirms the conjecture of Seiberg and Witten that the 
quantum moduli space of this theory is given by the double cover of the 
Atiyah-Hitchin manifold.

Instanton effects in the three-dimensional $N=2$ $SU(2)$ Yang-Mills 
theory were studied in \cite{3dN=2}, 
where it was shown that a superpotential is generated at the one-instanton 
level. Exact forms for the instanton-generated superpotential for 
$N=2$ theories coupled to matter were proposed in \cite{berkN=2,aha3d}. 
In Section 6,  we will test these proposals against explicit 
semiclassical calculations. The relevant formulae for constructing the 
instanton measure in the $N=2$ theory are given at the end of this section.

For $N=4$ SYM, in the absence of matter, the instanton zero modes 
were discussed in detail in \cite{dkmtv,dkm3d}. 
The Callias index theorem dictates that there 
are $4k$ bosonic and $4k$ fermionic zero 
modes, where $k$ is the instanton number. The classical action for the 
$k$-instanton is $S_k=kS_{\rm cl}$, where the single-instanton action,  
$S_{\rm cl}=8\pi^2M_W/e^2$ was defined in the previous section. 
The moduli space of instantons decomposes as 
$R^3\times (S^1\times {\tilde M}_k)/Z_k$. The $R^3$ and $S^1$ factors 
correspond to the spacetime position $X_{\mu}$, $\mu=0,1,2$, 
of the `centre of mass'  
and to the `centre of charge' $\theta$ of the instanton. 
We will refer to these four degrees of freedom as the `centre coordinates'.   
The relative moduli space ${\tilde M}_{k}$ is a smooth hyper-\Ka\ manifold 
of dimension $4(k-1)$ equipped with a natural metric\footnote{For the 
case $k=2$ it is a notable coincidence (and also 
a source of potential confusion) that the 
relative moduli space of two instantons ${\tilde M}_{2}$ is the 
double-cover of the AH manifold which is precisely the 
conjectured vacuum moduli space of the theory.} $\tilde{g_{ab}}$. 
Roughly speaking the coordinates on ${\tilde M}_{2}$, which we denote 
$Y^a;a=1,...,4(k-1)$, parametrize the relative separations and $U(1)$ charge angles of $k$ instantons. The discrete symmetry 
$Z_k$ acts on both the $S^1$ and on relative charges contained within 
${\tilde M}_k$. At leading semiclassical order, the path integral measure 
can be written as an integral over 
these collective coordinates with 
an appropriate Jacobian \cite{bern}. The additional prefactor which arises from including one-loop effects in the instanton 
background 
will be considered below. For the bosonic sector the  
collective coordinate measure consists of two factors corresponding to 
the centre and relative coordinates respectively. Explicitly \cite{dkm3d} we 
have ${\rm d}\mu_{B}={\rm d}{\bar \mu}_B{\rm d}{\tilde \mu}_B$ with, 
\begin{eqnarray}\label{bosmeas}
\int{\rm d}{\bar \m}_B=\frac{2^4\pi^2k^2M_W}{e^4}\int{\rm d}^3X\int_0^
{2\pi}{\rm d}\theta
&;&\int{\rm d}{\tilde \mu}_B=\frac{1}{(2\pi)^{2(k-1)}}\int \Pi_{a=1}^{4(k-1)} {\rm d}Y^a {\sqrt {\tilde g}} \ .
\end{eqnarray}
We have still to divide by the $Z_k$ symmetry factor. We will discuss
this in more detail when considering $k=2$ below.

In the $N=4$ theory, four of the total of $4k$ adjoint fermion zero modes 
arise from the action of four supersymmetry generators on the monopole 
background. The monopole background is invariant under the other 
SUSY generators.   Unlike the remaining fermionic zero modes, 
these modes are protected from lifting by SUSY. We introduce 
two two-component Grassmann collective coordinates $\xi_\a, \xi'_\a$. 
In the following we will need the large-distance behaviour of these modes. 
Consider a $k$-monopole solution with centre of mass coordinate, $X_{\mu}$. 
Then, for $|x-X| \gg  M_{W}^{-1}$, we have \cite{dkmtv,dkm3d}
\begin{eqnarray}
\lambda^{\rm LD}_{\alpha} & =  & 8\pi k
\left(S_{\rm F}(x-X)\right)_{\alpha}^{\ \beta}\xi_{\beta} \nonumber \\
\psi^{\rm LD}_{\alpha} & =  & 8\pi k
\left(S_{\rm F}(x-X)\right)_{\alpha}^{\ \beta}\xi'_{\beta}\ ,
\label{ld}
\end{eqnarray} 
where $S_{\rm F}(x)=\gamma^\mu x_\mu/(4\pi |x|^{2})$ is the three-dimensional 
Dirac fermion propagator. These modes are the SUSY partners of the 
centre coordinates, $X_{\mu}$ and $\theta$. As in \cite{dkm3d}, 
we also introduce Grassmann collective coordinates $\alpha^{a}$, 
$a=1,\ldots 4(k-1)$ which correspond to the remaining $4(k-1)$ adjoint 
fermion zero modes. These coordinates are Grassmann numbers rather than two-component Grassmann spinors, reflecting 
the fact that 
the number of adjoint zero modes is exactly 
half that of the $N=8$ case considered in \cite{dkm3d}. These modes are not 
protected by supersymmetry and we will see below that they are lifted in the 
presence of additional matter multiplets. 
As for the bosonic zero modes, it is convenient to write the 
measure for the adjoint fermionic collective coordinates as a product of
two factors corresponding to the centre and relative coordinates, 
$d\mu_{F}=d{\bar \mu}_{F}d{\tilde \mu}_{F}$ with,  
\begin{eqnarray}\label{fermeas}
\int\,{\rm d}{\bar \m}_F=\frac{e^4}{2^8k^2\pi^4M_W^2}\int\,{\rm d}^2\xi{\rm d}^2\xi'
&;&\int\,{\rm d}{\tilde \mu}_F=\int\, \Pi_{a=1}^{4(k-1)}{\rm d}\a^a \frac{1}{{\sqrt {\tilde g}}}\ ,
\end{eqnarray}
where ${\rm d}^2\xi=(1/2){\rm d}\xi_1{\rm d}\xi_2$.
 
So far, we have only considered the vector multiplets. We now include 
hypermultiplets, with masses as in \eqn{Sm}.
The relevant zero modes are solutions of the Dirac equation for the 
hypermultiplet fermions in the instanton background. 
The number of linearly independent solutions is 
given by the Callias index theorem \cite{cal}. Following Weinberg 
\cite{w1},  we define the operators,  
\begin{eqnarray}
\Delta^{R}_{-}(\vec{m})&=&D^2_{{\rm cl}}+2\g^\m\uBmu+|m|^2+({\tilde m}+
\uphi_3)^2 \nonumber\\
\Delta^{R}_{+}(\vec{m})&=&D^2_{{\rm cl}}+|m|^2+({\tilde m}+\uphi_3)^2\ ,
\label{massdeltas}
\end{eqnarray}
where where $\uB^\m_{\rm cl}=(1/2)\epsilon^{\m \n \rho }
\uv_{\n \rho }^{\rm cl}$ 
is the magnetic field of the BPS monopole and the superscript $R$ denotes the 
representation of the gauge group generators appearing in (\ref{massdeltas}). 
In the following $R=A$ denotes the adjoint representation and $R=F$ denotes 
the fundamental representation. 
The number of zero modes for a three-dimensional 
Dirac fermion transforming in the representation $R$ is 
given by the limit $\alpha^2\rightarrow 0$ of the regularized trace, 
\begin{equation}
{\cal L}_{\vec {m}}(\alpha^2)=\Tr \left[\frac{\alpha^2}{\Delta^{R}_{-}(\vec{m})+\a^2}- 
\frac{\a^2}{\Delta^{R}_{+}({\vec m})+\a^2}\right]\ .
\end{equation}
For zero complex mass, this trace was evaluated in  \cite{berkN=2}
(see appendix of the second reference). The generalization to non-zero 
complex mass is straightforward and yields,  
\begin{equation}
{\cal L}(\a^2)=\frac{\a^2}{\a^2+|m|^2}\sum_{w}\,\frac{(w^2 M_W+w{\tilde m})k}
{(\a^2+|m|^2+({\tilde m}+wM_W)^2)^{1/2}}\ ,
\label{indexd}
\end{equation}
where the trace over gauge indices has been exchanged for a sum 
over weights, $w$, of the representation $R$. Putting all masses to zero and 
specifying the adjoint representation, which has weights $1,0$ and $-1$, 
the number of zero modes 
is $2k$ for each species of Dirac fermion. As the $N=4$ vector multiplet 
contains two such species this result agrees with the counting of adjoint 
fermion zero modes given above. For each species of massless 
fermion in the fundamental representation (weights $\ft12$ and $-\ft12$) 
the number is reduced to $k$. 
In the presence of the complex mass all zero modes are lifted. 
However, in the presence of only a real mass, the zero mode structure 
depends on the relative values of ${\tilde m}$ and $M_W$. 
For $M_W>2|{\tilde m}|$ there 
are $k$ zero modes, and for $M_W<2|{\tilde m}|$ there are none \cite{cal}. 
For $M_W=2{\tilde m}$, the zero modes become non-renormalisable. 
This corresponds to the point on the Coulomb branch where the quarks 
are classically massless. 

Just as for the other modes, we can introduce collective coordinates for the 
hypermultiplet zero modes. We focus on fundamental hypermultiplets 
with all masses set to zero and hence each species of Dirac fermion 
has $k$ zero modes.     
For each hypermultiplet with Dirac fermions $\psi_{i}$ and 
${\tilde \psi}_{i}$, we can expand the zero mode solution of the 
Dirac equation in a complex orthonormal 
basis with Grassmann coefficients $\l^A, A=1,...,k$ 
\begin{equation}
\psi_i=\rho_i^A\l^A \qquad {\tilde \psi}_i={\tilde \rho}_i^A\l^A\ ,
\end{equation}
where $i=1,...,N_f$ and
\begin{equation}
\int\, {\rm d}^3x\, \l^{A\dagger}\l^B=\d^{AB}\ .
\end{equation}
Note that the charge conjugate fermions have no zero mode solutions, hence we set ${\bar \psi}_i={\bar {\tilde \psi}}_i=0$. 
Using this normalization,  the collective coordinate measure for the 
zero modes of $N_{f}$ fundamental hypermultiplets is
\begin{equation}\label{fundmeas}
\int {\rm d}\mu_{f}=\Big(\frac{e^2}{2\pi}\Big)^{kN_f}\int\, \Pi_{i=1}^{N_f} \Pi_{A=1}^{k}\,{\rm d}\rho_i^A{\rm d}{\tilde \rho}_i^A\ .
\end{equation}
As mentioned above, the four adjoint fermion zero modes 
parametrized by $\xi_{\alpha}$, $\xi'_{\alpha}$ are protected by 
supersymmetry and cannot be lifted. However this is not the case for 
either the remaining adjoint fermion zero modes parametrized by $\a^a$ or the 
hypermultiplet fermi zero modes with coordinates 
$\rho^A_i$, ${\tilde \rho}^A_i$. As the hypermultiplet and vector multiplet 
fermions have opposite charge under the unbroken R-symmetry denoted 
$U(1)_{N}$, Seiberg and Witten \cite{sw3} suggested that these modes can be 
lifted in pairs. In the following we will exhibit this lifting explicitly.  
In fact a similar effect is known to occur for the case of a single 
adjoint hypermultiplet, 
which corresponds to the $N=8$ model discussed in \cite{dkm3d}. 
In that case the lifting of modes was due to the presence of a SUSY- and 
$U(1)_{N}$-invariant Grassmann quadrilinear term in the action of the 
instanton. The quadrilinear term was found by dimensional 
reduction of the collective coordinate Lagrangian for the low-energy 
dynamics of BPS monopoles in $(3+1)$-dimensional ${\cal N}=4$ SUSY YM.    
This procedure yielded a quadrilinear term proportional to the Riemann tensor 
on the instanton moduli space. 

In the present case of fundamental hypermultiplets, 
a similar term can be deduced 
by dimensional reduction of the corresponding theory in four dimensions. 
The relevant collective coordinate Lagrangian for the low-energy dynamics of 
BPS monopoles in four-dimensional 
${\cal N}=2$ SQCD has been given in \cite{ssz,sweden,gh}. It contains a term 
bilinear in $\alpha^{a}$ and in $\rho^{A}_{i}$ (${\tilde\rho}^{A}_{i}$). 
After reducing to three dimensions, the resulting instanton action is
\footnote{To compare with the results in \cite{gh}, one must perform an 
$SU(2)_N$ rotation. This acts on the vector multiplet scalars, rotating the 
vev $\vhiggs_1$ into $\vhiggs_3$, and also on the hypermultiplet fermions. 
The net effect of this is that our $\rho^A$ and $\l^A$ are the same as in
\cite{gh}. This fixes the normalisation of the $F$ term in \eqn{Slift}.} 
\begin{equation}
\tilde{S}=kS_{\rm cl}-ik\s-\frac{2\pi}{e^2} \left( \frac{1}{4}F_{ab}^{AB}\a^a\a^b(\rho_i^A\rho_i^B+
{\tilde \rho}_i^A{\tilde \rho}_i^B)\right)\ .
\label{Slift}
\end{equation}
Here $F^{AB}_{ab}$ is the self-dual curvature tensor of a certain $O(k)$ 
bundle over the instanton moduli space. In the case $k=2$, 
which is treated in detail in the next section, explicit formulae for this 
tensor are given in \cite{Manschr}. From a
three-dimensional perspective, the quadrilinear 
term comes from the Yukawa 
terms in the  action. The field equations set the scalars of the vector and 
hypermultiplet to be quadratic in the fermionic collective coordinates, 
yielding Yukawa terms quartic in these coordinates. 

As in any semiclassical instanton calculation, to complete the 
specification of the measure we must also consider the contribution of 
non-zero modes. Often in supersymmetric gauge theories these contributions 
cancel between bose and fermi degrees of freedom. 
However, as explained in \cite{dkmtv}, 
this cancellation does not occur in three-dimensional theories 
with $N=4$ SUSY, because of the spectral 
asymmetry of the Dirac operator in a monopole background. In the case of $N=4$ 
SYM considered in \cite{dkmtv} the residual factor involves the 
ratio of the determinants of the operators 
$\Delta_{\pm}=\Delta^{A}_{\pm}(\vec{m}=0)$,  
\begin{equation}
R_{V}=\left[\frac{{\rm det}(\Delta_{+})}{{\rm det'}(\Delta_{-})}
\right]^{\frac{1}{2}}=(2M_W)^{2k}\ .
\label{detvec}
\end{equation} 
Here $\Delta_{+}$ is positive and has no zero modes, while $\Delta_{-}$ has $2k$ zero modes \cite{w1}. The prime in 
\eqn{detvec} denotes the removal of these zero modes. 

In the present case, the contribution of the vector multiplet is given by 
(\ref{detvec}), and there are additional one-loop factors in the measure for 
each hypermultiplet. 
Including scalar and fermion contributions from each fundamental 
hypermultiplet, the Gaussian 
integral over quadratic fluctuations around the instanton yields the factor 
\begin{equation}
R_{H}=\prod_{i=1}^{N_{f}}\left[\frac{{\rm det'}\Delta^{F}_{-}(\vec{m}_{i})}
{{\rm det}\Delta^{F}_{+}(\vec{m}_{i})}\right]^{\frac{1}{2}}\ .\label{Ddet}
\end{equation}
Using identical manipulations to those given in \cite{dkmtv}, 
each term in this product may be related to an integral over the regularized 
trace,  
(\ref{indexd}) 
\begin{equation}\label{formD}
R_{H}=\prod_{i=1}^{N_{f}} 
\left[\lim_{\alpha \rightarrow 0}
\Big(\alpha^{n_{i}}\exp[\int_\alpha^\infty \frac{{\rm d}\mu}{\mu}
{\cal L}_{\vec{m}_{i}}
(\mu)]\Big)^{-\frac{1}{2}} \right] \ ,
\end{equation}
where $n_{i}$ is the number of zero modes of the $i$'th hypermultiplet. 

For zero masses $\vec {m}_{i}=0$, 
the Callias index theorem tells us that $n_{i}=k$ and hence we have 
a one-loop factor 
\begin{equation}
R_{H}=(M_W)^{-kN_f}\ .\label{D0}
\end{equation}
In the next section, we consider one massless fundamental hypermultiplet 
with \eqn{D0} the relevant formula. 
On the other hand with all complex masses chosen to be non-zero, $n_{i}=0$ 
for each $i$ and we obtain, 
\begin{equation}\label{Dm}
R_{H}=\prod_{i=1}^{N_f}\left[\frac{\sqrt{|m_i|^2+({\tilde m}_i+\frac{1}{2}
M_W)^2}+{\tilde m}_i+\frac{1}{2}M_W}{\sqrt{|m_i|^2+({\tilde m}_i-\frac{1}{2}
M_W)^2}+{\tilde m}_i-\frac{1}{2}M_W}\right]^{-k/2}\ .
\end{equation} 
Notice that sending one of the masses to infinity reduces the corresponding 
factor to unity and the $N_f$ flavor theory flows to the 
$N_f-1$ flavor theory. 

As in \cite{PP,dkm3d}, one can also consider a 
theory with $N=8$ supersymmetry by adding a
 single massless adjoint hypermultiplet to $N=4$ SYM. In this case, the 
adjoint hypermultiplet yields a one-loop factor $R_{V}^{-1}$ and
cancels the determinant from the vector multiplet. Alternatively, 
one can consider mass deformed $N=8$ where the adjoint hypermultiplet 
is given a mass. For a non-zero complex mass, the ratio of determinants 
for a massive adjoint hypermultiplet is,
\begin{equation}\label{adjdet}
R_{A}=\left[\frac{\sqrt{|M|^2+({\tilde M}_i+M_W)^2}
+{\tilde M}_i+M_W}{\sqrt{|M|^2+({\tilde M}_i-M_W)^2}+
{\tilde M}_i-M_W}\right]^{-k}\ .
\end{equation}

Finally we will briefly state the modifications required to obtain the correct 
instanton measure in the $N=2$ theory with matter. 
The main difference is that the 
number of adjoint fermion zero modes in the instanton background is halved. 
In the $k$-instanton sector there are now $2k$ adjoint fermion zero modes 
of which only two are protected by supersymmetry. The corresponding 
collective coordinate measure is given by,     
$d\nu_{F}=d{\bar \nu}_{F}d{\tilde \nu}_{F}$ with,  
\begin{eqnarray}\label{fermeasN=2}
\int\,{\rm d}{\bar \nu}_F=
\frac{e^2}{2^4k\pi^2M_W}\int\,{\rm d}^2\xi
&;&\int\,{\rm d}{\tilde \nu}_F=\int\, \Pi_{a=1}^{2(k-1)}{\rm d}\a^a 
{\tilde g}^{-\frac{1}{4}}\ .
\end{eqnarray}
Similarly the resulting one-loop determinant factor differs from that of the 
$N=4$ theory by the contribution of the additional massless 
adjoint half-hypermultiplet which is present in the latter theory. Explicitly 
$R_{V}$ of equation (\ref{detvec}) is replaced by;
\begin{equation}\label{detvecN=2}
S_{V}=\left[\frac{{\rm det}(\Delta_+)}{{\rm det}'(\Delta_-)}\right]^{3/4}
=(2M_W)^{3k}\ .
\end{equation}  
It is also likely that the exact form of the 
Grassmann quadrilinear term in the multi-instanton action (\ref{Slift}) 
is different in the $N=2$ theory. However the form of this 
term is constrained by invariance under the $U(1)_{N}$ symmetry of 
the $N=2$ theory. As mentioned above,  zero modes of the 
vector multiplet fermions have opposite $U(1)_{N}$ charge to those of the 
hypermultiplet fermions. The multi-instanton action will therefore 
have the general form,    
\begin{equation}
\tilde{S}_{N=2}=kS_{\rm cl}-ik\s+O(\a^{2}\rho^{2})\ .
\label{SliftN=2}
\end{equation}
In fact this symmetry argument will be enough to show that there are 
no instanton corrections to the superpotential in the $N=2$ theory with 
massless hypermultiplets.  

It is possible to flow from the $N=4$ to the $N=2$ theory by adjusting 
the mass of the extra adjoint half-hypermultiplet. For non-zero complex mass, 
this augments the ratio of determinants \eqn{detvecN=2} by the 
factor $R_A^{1/2}$.

\section{Instanton Effects in $N=4$ Theories}
We will compute the leading order exponential corrections to the Riemann
tensor of the Coulomb branch metric (\ref{seff1}). When combined with
the 
perturbative result and the constraints of the 
hyper-K\"ahler condition this will 
prove sufficient to determine the metric fully. The Riemann tensor
appears in 
the low energy effective action 
(\ref{seff1}) as the coefficient of a four-fermion vertex. To determine
the 
instanton contribution to this vertex we therefore evaluate the
large-distance behaviour of the four-fermi correlator,  
\begin{equation}
G^{(4)}(x_1,x_2,x_3,x_4)=\langle\l_\a(x_1)\l_\b(x_2)\psi_\g(x_3)\psi_\delta(x_4)\rangle \
,\label{4fermi}
\end{equation}
where the fermion fields take their zero-mode values in the 
instanton background. 
Collecting the various factors from the previous section the leading 
semiclassical contribution to this correlator is, 
\begin{equation}  
G^{(4)}(x_1,x_2,x_3,x_4)=\int {\rm d}\mu_{B}{\rm d}\mu_{F}{\rm d}\mu_{f}
\l^{{\rm LD}}_\a(x_1)\l^{{\rm LD}}_\b(x_2)\psi^{{\rm LD}}_\g(x_3)\psi^{{\rm LD}}_\delta(x_4) 
\, R_{V}R_{H}\exp\left(-\tilde{S}\right)\, .
\label{fourf}
\end{equation}
Definitions of 
the various ingredients in this formula can be found in equations 
(\ref{bosmeas}), (\ref{ld}), (\ref{fermeas}), (\ref{fundmeas}), 
(\ref{Slift}), (\ref{detvec}) and (\ref{D0}). 

For (\ref{fourf}) to yield a non-zero answer it is necessary to saturate
each of the Grassmann integrations appearing in the fermionic measure 
${\rm d}\mu_{F}{\rm d}\mu_{f}$. The integrals over the SUSY coordinates
$\xi_{\alpha}$ 
and $\xi'_{\alpha}$ are saturated by the explicit insertion of the 
four fermion fields as given in (\ref{ld}). In the $k$ instanton sector,
there are also Grassmann integrations corresponding to the $4(k-1)$
remaining  
vector multiplet fermion zero modes and $2kN_{f}$ fermion zero modes
from the 
hypermultiplets.  These integrations can only be saturated by bringing
down 
powers of the Grassmann quadrilinear term appearing in $\tilde{S}$.
Clearly 
this only gives a non-zero result if the number of remaining zero
modes 
from the vector and hyper-multiplets are equal. This requires that
\begin{equation}
4(k-1)=2kN_{f}\, .
\label{xcond}
\end{equation}          
Thus we see that different instanton sectors contribute to the 
metric on the Coulomb branch for different values of $N_{f}$ \cite{sw3}.
For $N_{f}=0$, the above condition is trivially satisfied for $k=1$.
This 
corresponds to the one-instanton contribution in $N=4$ SYM theory 
which was calculated in \cite{dkmtv}. Similarly, for a single massless
hypermultipet, (\ref{xcond}) requires $k=2$, 
which corresponds to a non-vanishing two instanton contribution that we
will calculate below. For $N_{f}>1$ there are no instanton corrections
to the 
metric on the Coulomb branch.     

We now specialize to the case $N_f=1$, $k=2$ where the remaining 
Grassmann integrals are saturated by bringing down two powers of the 
quadrilinear term in $\tilde{S}$ which gives, 
\begin{eqnarray}\label{G4}
&&G^{(4)}(x_{1},x_{2},x_{3},x_{4})= 
2^{15}\pi^{3}M_{W}\Big(\frac{1}{8\pi^2}\int_{M_{k}}
\,F\wedge F\Big)\exp(-2S_{\rm cl}+2i\sigma)\times \\
&&\hspace{1cm}\int{\rm d}^3X\, 
\epsilon^{\alpha'\beta'}S_{\rm F}(x_{1}-X)_{\alpha\alpha'}
S_{\rm F}(x_{2}-X)_{\beta\beta'}
\epsilon^{\gamma'\delta'}S_{\rm F}(x_{3}-X)_{\gamma\gamma'}
S_{\rm F}(x_{1}-X)_{\delta\delta'}\nonumber \ ,
\label{corr3}
\end{eqnarray}
where $F$ is a two-form on the relative moduli space of two instantons
$M_{2}$ which is constructed from the components of 
curvature tensor of the $O(2)$ bundle which appears in (\ref{Slift}). 
Choosing coordinates $Y^{a}$ on $M_{2}$, with $a=1,2,3,4$, we
have,   
\begin{equation}
F\equiv \ft12 F_{ab}^{(12)}{\rm d}Y^a\wedge {\rm d}Y^b\ . 
\label{fst}
\end{equation}
The moduli space in question is exactly the AH manifold. We must also 
include the effect of the $Z_2$ symmetry acting on the monopole moduli 
space. It's action on the centre of charge 
is $\theta\rightarrow\theta +\pi$. 
It simultaneously acts on the $\psi$ coordinate on the AH manifold 
(see \eqn{1forms} below) as $\psi\rightarrow \psi+\pi$ . 
As $\theta$ does not appear in the integrand, we use it to restrict the 
range of $\psi$ to $2\pi$. Fortunately the integral of $F\wedge F$ 
over the AH manifold has been 
evaluated explicitly in \cite{ssz,gh} where this integral 
appears as part of the volume contribution to the index of the Dirac
operator. Explicitly, we find,
\begin{equation}
\frac{1}{8\pi^{2}}\int_{M_{2}}\, F \wedge F = \frac{1}{8}\ .
\label{ghintegral}
\end{equation}    
The resulting four-point function corresponds to an instanton-induced 
vertex in the low-energy effective lagrangian  
of the form 
\begin{equation}
{\cal L}_{4F}=\kappa {\bar \l}^2{\bar \psi}^2\exp \left( -2S_{\rm
cl}+2i\s \right)\ ,\label{L4F}
\end{equation}
with 
\begin{equation}
\kappa=2^{10}\pi ^3 M_W \left(\frac{2\pi}{e^2}\right)^4\ .\label{kappa}
\end{equation}
The four powers of $2\pi/e^2$ reflect the normalisation of the fermion
kinetic terms.

All that remains is to compare the result of our calculation with the 
prediction of Seiberg and Witten \cite{sw3}. We will be brief as 
the comparison is almost identical to that performed for the 
$N_{f}=0$ case in \cite{dkmtv}. 
As explained in section 3, $N=4$ SUSY dictates that the exact 
effective action has the form (\ref{seff1}). Hence, to obtain an 
exact solution of the low-energy theory one must specify the hyper-\Ka\
metric on the target space ${\cal M}$. Seiberg and Witten have 
conjectured that ${\cal M}$ is the double cover of the AH manifold 
which we have already met in an apparently unrelated context 
as the moduli space of two instantons. Hence the relevant metric is the
one 
constructed explicitly by Atiyah and Hitchin and described in detail in
\cite{AH}. 

The Atiyah-Hitchin manifold admits an $SO(3)$ isometry and hence the 
metric can  be written in the form \cite{gp}
\begin{equation}
g_{ab} {\rm d}X^{a} {\rm d}X^{b}=f^{2}(r){\rm d}r^2+a^{2}(r)\sigma_{1}^{2}+
b^{2}(r)\sigma_{2}^{2}+ c^{2}(r)\sigma_{3}^{2}\ .    
\label{metric}
\end{equation}
Here, $\s_i$, $i=1,2,3$ are the three 
left-invariant one forms on the $SO(3)$ orbit parametrized by 
Euler angles $\theta$, $\phi$ and $\psi$ with ranges,  
$0\leq \theta \leq \pi$, $0\leq \phi <2\pi$ and  $0\leq \psi<2\pi$, 
\begin{eqnarray}
\sigma_{1} & =&-\sin\psi \ {\rm d}\theta+\cos\psi\sin\theta \ {\rm d}\phi 
\nonumber\\     
\sigma_{2} &= &\cos\psi \ {\rm d}\theta+\sin\psi\sin\theta \ {\rm d}\phi 
\nonumber\\   
\sigma_{3} &= &{\rm d}\psi+\cos\theta \ {\rm d}\phi \, .      
\label{1forms}
\end{eqnarray}
Its double cover has an $SU(2)$ isometry with $0<\psi<4\pi$, the
functions $a,b$ and $c$ are still the same.
It is convenient to define cartesian coordinates for the three
non-compact 
directions, 
\begin{equation}
X=r\sin\theta\cos\phi \qquad
Y=r\sin\theta\sin\phi \qquad
Z=r\cos\theta  \ ,
\label{cart}
\end{equation}  
and also to define complex coordinates; 
\begin{eqnarray}
z_{1}=\frac{1}{\sqrt{2}}(X-iY) & \qquad{} \qquad{} \qquad{} & 
z_{2}=\frac{1}{\sqrt{2}}(Z- i\psi)\ .
\label{cbasis}
\end{eqnarray}
In this basis, one can show that the leading exponential correction in
the Riemann tensor only appears in
$R_{1212}$ and its complex conjugate \cite{dkmtv}.
The function $f(r)$ depends on the exact definition of the radial
parameter
$r$. Following \cite{gm}, we choose $f=-b/r$
\footnote{This approach differs slightly from \cite{dkmtv} where 
the relationship $r=S_{\rm cl}$ was taken to define the radial 
coordinate and hence $f$. Fixing $f=-b/r$ means that the 
relationship between $r$ and $S_{\rm cl}$ will receive quantum 
corrections.}. 
The hyper-\Ka\ condition forces 
the remaining components $a(r)$ $b(r)$ and $c(r)$ to obey a set of three
non-linear ODE's which were analysed in \cite{AH}. 
After identifying the appropriate 
boundary conditions, explicit forms can be found for 
$a(r)$, $b(r)$ and $c(r)$ in terms of elliptic functions. For the
present 
purposes we only require the large distance asymptotic forms of these 
components,    
\begin{equation}\label{asmptmetr}
a^2=r^2(1-\frac{2}{r})-4r^2e^{-r}+... \quad b^2=r^2(1-\frac{2}{r})+
4r^2e^{-r}+...\quad c^2=4(1-\frac{2}{r})^{-1} +...\ .
\end{equation}
Further corrections are supressed by powers of $1/r$ or $\exp (-r)$. 

At this point we can substitute (\ref{metric}), 
with the above asymptotic forms for the components $a$, $b$ and $c$, for
the 
metric in the effective action (\ref{seff1}). In particular, we can
compare 
the power-law terms with the results of our perturbative calculation. 
This comparison yields the one-loop identifications, 
$(X,Y,Z)=(2S_{\rm cl}/M_{W})(\phi_{1},\phi_{2},\phi_{3})$ and
$\psi=\sigma$. 
The first equality implies that $r=2S_{\rm cl}+O(1/S_{\rm cl})$ 
(see footnote below). 
In the following it is important that this relation is correct to the order 
shown. In particular, one may verify explicitly that the additive constant 
which could in principle appear at one-loop on the RHS of the relation 
vanishes.  At this order, the metric can be
written in the form \eqn{asmetric} with $U=1-2/r$.
The second equality implies that we are indeed dealing with the 
double cover instead of the single cover of AH (in which case we had 
$\psi=\s/2$ \cite{dkmtv}). 

 A similar comparison
of the fermion kinetic term for the fermions $\O^{a}$ in (\ref{seff1})
with its counterpart in the one-loop effective action allows us to make
the 
following identifications in the complex basis 
\begin{equation}    
\O^{1}= \frac{2S_{\rm cl}}{M_W}\bar \lambda \qquad \O^{2}= \frac{2S_{\rm
cl}}{M_W}\bar \psi \ , 
\label{fident}
\end{equation}
and similar for the conjugated fermions. 
With the above identifications, the exact effective action gives rise to
a 
vertex which, just like (\ref{L4F}), 
couples four fermions of the same chirality:  
\begin{equation}
{\cal L}_{4F}=\frac{1}{4} K\left( \frac{2S_{\rm
cl}}{M_W}\right)^4R_{1212}{\bar \l}^2{\bar \psi}^2+\mbox{h.c.}\ .
\label{result}
\end{equation}
Hence, to complete the comparison, we must use the explicit 
asymptotic form (\ref{asmptmetr}) of the metric to compute the leading
exponentially suppressed contribution to $R_{1212}$. The relevant
component  
of the Riemann tensor was computed to the required order in
\cite{dkmtv}. Using the relation between the coordinates and 
the massless fields up to one-loop, we have 
\begin{equation}
R_{1212}=8S_{{\rm cl}}e^{-2S_{{\rm cl}}+2i\sigma}\ .
\label{Rtensor}
\end{equation}
Substituting the above result for $R_{1212}$ in 
(\ref{result}) we find precise agreement with the instanton induced
vertex 
(\ref{L4F}). 
The conjugated term will be generated by the corresponding
two-anti-instanton process.
This confirms the conjecture of Seiberg and Witten that the
quantum Coulomb branch of the theory is the double cover of the 
Atiyah-Hitchin manifold. 

In fact, as in \cite{dkmtv}, the result of our 
instanton calculation is actually sufficient to deduce this
correspondence 
from first principles. That the exact Coulomb branch 
metric has the form (\ref{metric}) is a consequence of the $SU(2)_{N}$
$R$-symmetry of the theory. As mentioned above, the 
hyper-\Ka\ condition, which is necessary for $N=4$ SUSY, leads to a set
non-linear ODE's for the metric components $a(r)$, $b(r)$ and $c(r)$. A
simple analysis of these equations given in \cite{dkmtv} shows that
one-loop and 
two-instanton data are enough to fix the boundary conditions and specify
a unique solution.    

In \cite{sw3}, Seiberg and Witten also proposed 
that introducing a mass for the hypermultiplet in this theory 
corresponds to Dancer's deformation of the double-cover of AH manifold.
As we 
discussed in Section 3, our perturbative results agree with this
proposal. 
Unfortunately, there are no explicit formulae available for the 
the exponentially suppressed corrections to the asymptotic metric on the
deformed manifold and thus it is not possible to perform a
non-perturbative 
check in the massive case. On the other hand, we could instead 
assume that the proposal is correct 
and use the instanton techniques to predict the metric, along the 
lines of \cite{cd}. 

There is one feature of the massive theory is which is 
straightforward to check: the RG flow to the $N=4$ 
theory without matter as the hypermultiplet mass is taken to infinity.
Briefly, the mass term lifts the hypermultiplet zero modes and therefore
allows a non-zero one-instanton contribution. Mass dependence is also 
introduced in the one-loop prefactor $R_{H}$ as per equation (\ref{D0}).
One may easily check that, in the decoupling limit, this 
contribution reproduces the 
one-instanton effect of the $N_{F}=0$ theory which 
was calculated in \cite{dkmtv}.  
   
\section{Instanton Effects in $N=2$ Theories}

We now turn to the theory with $N=2$ supersymmetry. We will initially 
specialize to the case of $N_{f}$ massless hypermultiplets in the fundamental 
representation.  As discussed in 
Section 3, $N=2$ supersymmetry allows the generation of a superpotential 
which leads to fermion bilinear terms in the effective action. To determine 
the instanton contribution to the superpotential we therefore calculate 
the large-distance behaviour of the the two-point correlator,    
\begin{equation}\label{G2}
G^{(2)}(x_1,x_2)=<\lambda_\alpha(x_1)\lambda_\beta(x_2)>\ .
\end{equation}
Collecting all the relevant factors, the $k$-instanton contribution to this 
correlator can be written as 
\begin{equation}  
G^{(2)}(x_1,x_2)=\int\, {\rm d}\mu_{B}{\rm d}\nu_{F}{\rm d}\mu_{f}\,
\l^{{\rm LD}}_\a(x_1)\l^{{\rm LD}}_\b(x_2)
\, S_{V}R_{H}\exp\left(-\tilde{S}_{N=2}\right)\ , 
\label{twof}
\end{equation}
where the various quantities appearing in the above expression 
are defined in equations 
(\ref{bosmeas}), (\ref{ld}), (\ref{fermeasN=2}), (\ref{fundmeas}), 
(\ref{SliftN=2}), (\ref{detvecN=2}) and (\ref{D0}). 

To obtain a non-zero contribution, all the Grassmann integrations 
appearing in the fermionic part of the measure must be saturated. 
As in the $N=4$ case, the modes which correspond to supersymmetry 
transformations of the instanton are saturated by the massless fermion fields 
given in (\ref{ld}). The remaining fermion zero modes comprise 
$2(k-1)$ modes from the vector multiplet each with $U(1)_{N}$ charge $+1$ and 
$2kN_{f}$ modes from the massless hypermultiplets with charge $-1$. As these 
modes can only be lifted in $U(1)_{N}$ neutral pairs, the condition for a 
non-zero contribution is, 
\begin{equation}
2(k-1)=2kN_{f}
\ .
\label{condN=2}
\end{equation}
Clearly this condition can only be satisfied for $k=1$ and $N_f=0$.
Thus we will calculate a one instanton effect in the $N=2$ without matter 
couplings.    
    
Performing the integration over bosonic and fermionic zero modes of 
the one instanton solution, we determine
\begin{equation}\label{n=2}
G^{(2)}(x_1,x_2)=\frac{2^9\pi^3}{e^2}M_{W}^3{\rm exp}(-S_{\rm cl}+i\sigma)
\int {\rm d}^3x\epsilon^{\alpha '\beta '}S_{{\rm F}\alpha\alpha '}
(x_1-x)S_{{\rm F}\beta \beta '}(x_2-x)\ . 
\end{equation}
This correlator corresponds to a
two-point vertex in the low-energy effective action of exactly the type 
expected from the general expression (\ref{superp}). By comparing the 
fermion kinetic term for the massless fermion $\lambda$ with the kinetic 
term of the fermion $\Psi$ in (\ref{comp}) one finds that these fermions 
are related as $\lambda=(M_W/S_{\rm cl})\bar{\Psi}$. Comparing the 
contribution of the fermion bilinear vertex in (\ref{superp}) to $G^{(2)}$ 
with (\ref{n=2}) we can determine the superpotential,    
\begin{equation}\label{superpot}
W(Z)=\frac{e^4S_{\rm cl}^3}{2^4\pi^5}\exp (-S_{\rm cl}+i\sigma)=
\frac{e^2}{2^4\pi^5}\exp(-Z)\ ,
\end{equation}
where $Z$ is the complex coordinate on the Coulomb branch given at 
one-loop by (\ref{Z}). This is the superpotential first obtained by 
Affleck, Harvey and Witten in \cite{3dN=2} although the calculation of 
one-loop fluctuations around the instanton background included above is new 
as is the overall normalisation of the superpotential. 
In particular, the prefactor of $S_{\rm cl}^{3}$ in 
(\ref{superpot}) which comes from the one-loop determinants precisely reflects 
the fact that $W$ is holomorphic with respect to the one-loop corrected 
complex structure. 

In the case of massive hypermultiplets the story is 
more complicated. The complex mass term explicitly breaks the $U(1)_{N}$ 
symmetry and hence the condition (\ref{condN=2}) for a non-zero contribution 
no longer applies. In particular, when each complex mass is non-zero, 
there are one-instanton contributions to the 
superpotential for each value of $N_{f}$. As all the hypermultiplet zero 
modes are lifted, we must omit the factor $d\mu_{f}$ in (\ref{twof}).  
In the massive case we must also use 
the definition (\ref{Dm}) for the one-loop factor $R_{H}$. The resulting 
superpotential is:
\begin{eqnarray}      
\label{superpotm}
W(Z)&=&\frac{e^4S_{\rm cl}^{3}}{2^4\pi^5}\left(
\prod_{i=1}^{N_f}\left[\frac{\sqrt{|m_i|^2+({\tilde m}_i+\frac{1}{2}M_W)^2}
+{\tilde m}_i+\frac{1}{2}M_W}{\sqrt{|m_i|^2+({\tilde m}_i-\frac{1}{2}M_W)^2}
+{\tilde m}_i-\frac{1}{2}M_W}\right]\right)^{-\frac{1}{2}}
\exp (-S_{\rm cl}+i\sigma)\nonumber\\
&=&\frac{e^4}{2^4\pi^5}\exp(-Z) \ , 
\end{eqnarray}
where we have used (\ref{Z1loop}) to express W as a holomorphic function of 
the complex field $Z$. This result agrees with the proposal for the exact 
superpotential made in \cite{berkN=2} up to the overall normalisation 
which is not explicitly specified in that reference. 
When expressed as a function of $Z$, we see that the one-instanton 
contribution to the superpotential has no dependence on the number of 
flavors or their masses
\footnote{If $m_i=0$ and $M_W>2|\tilde{m}_i|$, the change of variables 
between $Z$ and $S_{\rm cl}$ is ill-defined. In this case the 
superpotential vanishes.}. 
This dependence has been completely absorbed in 
the definition of $Z$, or, in other words, in the complex structure. 
In principle there can also be multi-instanton corrections in the 
massive case, however we will not attempt to calculate them here.   

Finally, we note that we may flow from the $N=4$ to the $N=2$ theory 
by the addition of a massive half-hypermultiplet in the adjoint 
representation. As in the fundamental case, in order to lift 
the extra fermionic adjoint zero modes, we require either $M>0$ 
or $M_W<|\tilde{M}|$. The $k$-instanton contribtution to the two fermi 
correlator, $G^{(2)}$, is equal to the $k$-instanton 
contibution to the four-fermi correlator, $G^{(4)}$ in the theory with 
full $N=4$ supersymmetry. More precisely, the prefactor to $G^{(2)}$ is 
equal to that of $G^{(4)}$ multiplied by the determinant \eqn{adjdet} 
and the factor $M(2M_W/M)^k$. This combination has the right behaviour in the 
decoupling limit $M\rightarrow\infty$. 

\section*{Acknowledgements}

We are grateful to J. de Boer, C. Houghton, V. Khoze, N. Manton, 
M. P. Mattis, A. Mountain and P. Sutcliffe for useful discussions. 
DT is grateful to the University of Washington for hospitality while 
this work was completed. The work of DT and SV was supported by PPARC.

\section*{Appendix A: Wilsonian Effective Action at One Loop}

In this appendix we calculate the one-loop contribution to the 
renormalisation of the coupling constant. The calculation is a 
direct generalisation of that performed in 
Appendix B of \cite{dkmtv} and employs the background field method. 
We restrict ourselves initially to $N=4$ multiplets, discussing the 
$N=2$ case at the end. Representations of the $N=4$ supersymmetry algebra 
allow for either vector or hypermultiplets. The vector 
multiplet is always in the adjoint representation of the $SU(2)$ gauge 
group, while hypermulitplets may be in either the adjoint or 
fundamental representation. Hypermultiplets may 
also have an arbitrary mass, ${\vec m}$, (see section 2). Integrating 
out the high momentum modes for any multiplet results in a term 
for the Wilsonian effective action containing the massless bosonic 
fields of the vector multiplet. This term is of the form
\begin{equation}
\ft12 C\,\int\,\frac{{\rm d}^3k}{(2\pi)^3}\{ A_\mu (-k) A_\nu (k)(
k^2g^{\mu\nu}-k^\mu k^\nu) + (A^\dagger (-k) A(k) +\phi(-k)\phi(k))k^2
\}\ ,
\end{equation}
where we have dropped terms of ${\cal O}(k^4)$. $C$ can be written in 
the general form
\begin{equation}
C=2\sum_w \int\frac{{\rm d}^3p}{(2\pi)^3}\frac{w^2\chi}
{(p^2-|{\vec m}+\sqrt{2}w{\vec\vhiggs}|^2)^2} 
=\frac{1}{4\pi}\sum_w\frac{w^2\chi}{|{\vec m}+\sqrt{2}w{\vec\vhiggs}|}
\ ,
\end{equation}
where $\chi=-1$ for a vector multiplet and $\chi =+1$ for a hypermultiplet 
and $w$ are the weights of the representation; $w= -1/2,+1/2$ 
for the fundamental representation and $w=-1,0,1$ for the adjoint. 
Comparing to the tree-level low-energy effective action, we find a 
finite renormalisation to the coupling constant. For the $N=4$ theory 
with one massless vector multiplet, $N_f$ fundamental 
hypermultiplets with mass ${\vec m}_i$, $i=1,...,N_f$ and $N_a$ adjoint 
hypermultiplets with mass ${\vec M}_I$, $I=1,...,N_a$, the renormalisation is 
\begin{eqnarray}\label{couprenorm}
\frac{2\pi}{e^2}&\rightarrow&\frac{2\pi}{e^2}-\frac{1}{2\pi M_W}
+\frac{1}{2^4\pi}\sum_{i=1}^{N_f}\left( |{\vec m}_i+{\vec\vhiggs}/\sqrt{2}
|^{-1}+|{\vec m}_i-{\vec\vhiggs}/\sqrt{2}|^{-1}\right) \nonumber \\
&&\hspace{1cm}+\frac{1}{4\pi}\sum_{I=1}^{N_a}\left(
|{\vec M}_I+\sqrt{2}{\vec\vhiggs}|^{-1}+|{\vec M}_I-
\sqrt{2}{\vec\vhiggs}|^{-1}\right)\ .
\end{eqnarray}
We mention a few special cases. For $N_a=0$ and ${\vec m_i}=0$, the 
behaviour of the renormalised coupling constant is ``$2-N_f$''. This 
contrasts the ``$4-N_f$'' behaviour in four-dimensional 
${\cal N}=2$ $SU(2)$ super Yang-Mills theory. 
In \cite{sw3} this was explained using an anomaly argument and 
noticing that in the background of the appropriate instanton the 
four-dimensional vector mutiplet fermions have twice as many zero modes 
as their three-dimensional counterparts due to extra super-conformal modes. 
See \cite{thompson} for a related discussion.

The theory with $N=8$ supersymmetry is obtained by taking $N_f=0$, $N_a=1$ 
with ${\vec M}=0$. As expected, the coupling constant is not renormalised 
in this case. 

Finally, we may view the $N=4$ theory as the $N=2$ theory with an 
additional adjoint half-hypermulitplet. Turning this around, the renormalised 
coupling constant for the pure $N=2$ theory is obtained from the $N=4$ theory 
by ``subtracting'' a massless adjoint half-hypermultplet. Formally 
inserting $N_f=0$ and $N_a=-1/2$, ${\vec M}=0$ into equation \eqn{couprenorm} 
yields
\begin{equation}\label{renccN=2}
\frac{2\pi}{e^2}\rightarrow\frac{2\pi}{e^2}-\frac{3}{4\pi M_W}\ ,
\end{equation}
which is indeed the correct renormalisation of the pure $N=2$ coupling 
constant. We may further add adjoint or fundamental half-hypemultiplets 
to this theory by augmenting equation \eqn{renccN=2} with the last 
two terms of \eqn{couprenorm}, resulting in a ``$3-N_f$'' behaviour 
for the renormalised coupling constant.

\end{document}